\documentclass[reprint,amsmath,amssymb,aps,showpacs]{revtex4-1}
\usepackage{graphicx}
\usepackage{dcolumn}
\usepackage{bm}
\usepackage{amsmath}
\usepackage{hyperref}
\usepackage{braket}
\usepackage{epstopdf}
\usepackage[sort&compress]{natbib}
\begin{document}
\title{Adiabatic transfer of energy fluctuations between membranes inside an optical cavity}
\author{Devender Garg}
\author{Anil K. Chauhan}
\author{Asoka Biswas}%
\email{abiswas@iitrpr.ac.in}
\affiliation{%
Department of Physics, Indian Institute of Technology Ropar, Rupnagar, Punjab 140001, India
}%
\date{\today}%

\begin{abstract}
A scheme is presented for the adiabatic transfer of average fluctuations in the phonon number between two membranes in an optical cavity. We show that by driving the cavity modes with external time-delayed pulses, one can obtain an effect analogous to stimulated Raman adiabatic passage (STIRAP) in the atomic systems. 
The adiabatic transfer of fluctuations from one membrane to the other is attained through a `dark' mode, that is robust against decay of the mediating cavity mode. The results are supported with analytical and numerical calculations with experimentally feasible parameters. 
\end{abstract}
\maketitle

\section{Introduction} \label{INTRO}
Controlling the quantum systems has been the main focus of research in various branches in physics and chemistry. The main goal in quantum control techniques is to develop  systematic methods for the active manipulation and control of quantum systems, to obtain a deterministic output.  It has witnessed many exciting applications, including coherent control of different molecular processes, e.g, photoassociation, photodissociation, and scattering \cite{Warren,Brumer}, quantum computing \cite{chuang}, and control of decoherence \cite{dd}. 

Most control schemes for quantum systems rely on its interaction with light. The optomechanical system \cite{Aspelmeyer} poses a suitable platform to explore such control techniques. 
In such systems, the light field inside an optical cavity and a mesoscopic mechanical oscillator, with a frequency far from the optical domain, interact with each other  through radiation pressure force. So far, several nontrivial quantum phenomena have been realized in optomechanical systems. This includes side band \cite{Marquardt2007,Teufel2011}  and near-ground-state  cooling \cite{Chan2011}, and squeezing \cite{squeeze} of a mechanical oscillator. The strong coupling \cite{Simon2009,Teufel2011circuit}  and quantum coherent coupling \cite{Verhagen2012} between cavity field mode and mechanical oscillator have also been achieved. This leads to potential applications of such systems into quantum communication and quantum information processing
\cite{Stannigel2012,njp,Rogers2014}. For example, a mechanical oscillator can mediate high fidelity state transfer between two optical cavities \cite{Narayanan2013}. These oscillators have been used to store optical information as a mechanical excitation \cite{Fiore2011}, as well. The state of a cavity field can even be coherently transferred into, stored-in and retrieved back from a mechanical oscillator \cite{njp,Palomaki2013}, allowing these oscillators to pose as quantum memory. 
In this paper, we proceed further to exploit the quantum aspects of the oscillators. Precisely speaking, we show how the quantum fluctuations can be coherently and deterministically transferred from one mechanical oscillator to the other. This clearly opens up an avenue of quantum communication between two truly mesoscopic systems.

\begin{figure}[h]
\centering
\includegraphics[width=6cm, height=3cm]{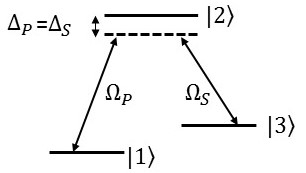}
\caption{Three-level $\lambda$ system}\label{lambda}
\end{figure}

The main result of this paper rely on stimulated Raman adiabatic passage (STIRAP) \cite{hioe,stirap} -  a quantum control technique to efficiently transfer population between two discrete quantum states (which are not dipole-coupled) of an atom,
 adiabatically using two resonant pulses. In a three-level $\Lambda$ configuration (see Fig. \ref{lambda}), one applies the pulses in a so-called counter-intuitive sequence, such that the pump pulse (with Rabi frequency $\Omega_P$) follows the Stokes pulse (with Rabi frequency $\Omega_S$) so as to evoke the population transfer from the state $|1\rangle$ to $|3\rangle$, without populating the intermediate excited state $|2\rangle$.
This process can be explained in terms of the eigenstates of the relevant Hamiltonian in interaction picture:
 \begin{equation}\label{3level}
 H=\frac{\hslash}{2}\begin{bmatrix}0 & \Omega_{P} & 0\\ \Omega_{P} & 2\Delta_{P}&\Omega_{S}\\0 & \Omega_{S} & 2\left(\Delta_{P}-\Delta_{S}\right) \end{bmatrix}
\end{equation}
where $\Delta_{P}$ and $\Delta_{S}$ are the single-photon detunings of the pump and the Stokes pulse, respectively. In two-photon (Raman) resonance $\Delta_{P}=\Delta_{S}$, the above Hamiltonian exhibits zero eigenvalue with the corresponding eigenstate
\begin{equation}\label{adiab}
|\psi_{D}\rangle \propto \Omega_S |1\rangle - \Omega_P |3\rangle.
\end{equation}
Suitable initial condition and time-dependence of the pulses ensures the formation of the above dark state. During subsequent time-evolution of the pulses, the population remains trapped in the dark state and gets deterministically and adiabatically transferred from the state $|1\rangle$ to $|3\rangle$. Note that the dark state $|\psi_D\rangle$ does not have any projection on the excited state $|2\rangle$ and therefore the evolution remains immune to the radiative decay. The STIRAP is robust against fluctuations of the parameters of the applied pulses, namely, the laser intensity, pulse timing and pulse shape \cite{stirap}.


In this paper, we describe an adiabatic process, akin to STIRAP, in the context of quantum optomechanics, in which two membranes are suspended inside an optical cavity. 
Note that, as mentioned above, while most of the research on optomechanical systems has confined itself into study of, e.g., cooling, squeezing, and realization of strong coupling (for a detailed review, see \cite{Aspelmeyer}), only recently, there have been growing interest on the adiabatic control of such systems.  The state transfer between two cavities of different wavelengths has been reported using the mediating mechanical mode. Tian \cite{Tian(2012)}, and Wang and Clerk \cite{Wang(2012)} have proposed simultaneously and independently that  it is possible to adiabatically transfer the quantum state between an optical cavity and a microwave cavity, using a mechanical resonator. In these proposals, one makes use of a mechanical dark mode, which is a superposition of the optical and the microwave mode and immune to mechanical dissipation so to allow the transfer of the intracavity field with high fidelity, without populating the mechanical mode. Later, Dong and his coworkers have experimentally demonstrated an adiabatic transfer of optical fields between two optical modes of a silica resonator, using such a mechanical dark mode \cite{Dong(2012)}, in the regime of weak optomechanical coupling. This further opens up the possibility of using optomechanical coupling into various applications without cooling the mechanical oscillator to its ground state.  Note that it is also possible to convert the optical wavelength from one cavity mode to the other, using their common radiation pressure coupling to the same mechanical mode  \cite{painter2012}, while a sequence of $\pi/2$-pulses can be used to transfer the states of the cavity mode to the mechanical mode  via their beam-splitter-type interaction \cite{Tian2010}.

Here we propose a technique to adiabatically transfer the energy fluctuations between two membranes suspended inside a cavity, that contrary to the previous reported works where such transfer occurs between two cavity modes.  Specifically, in absence of decay of the cavity modes and the membranes, one obtains a zero-eigenvalue adiabatic state, the time-evolution of which governs this transfer. We show that by choosing a suitable time-sequence of the external fields that drive the cavity modes, one can effectively make the optomechanical coupling time-dependent, so as to mimic a STIRAP-like process. We emphasize that during the adiabatic transfer, the decay of the cavity modes has negligible effects.

The paper is organized as follows. In Sec. \ref{MODEL}, we present the model and derive the effective Hamiltonian. We analyze the eigenstructure of this Hamiltonian and present numerical results in support of out analysis, in Sec. \ref{EAEV}. In Sec. \ref{con}, we conclude our paper.

\section{Model}   \label{MODEL}
\begin{figure}[h]
\centering
\includegraphics[width=.4\textwidth]{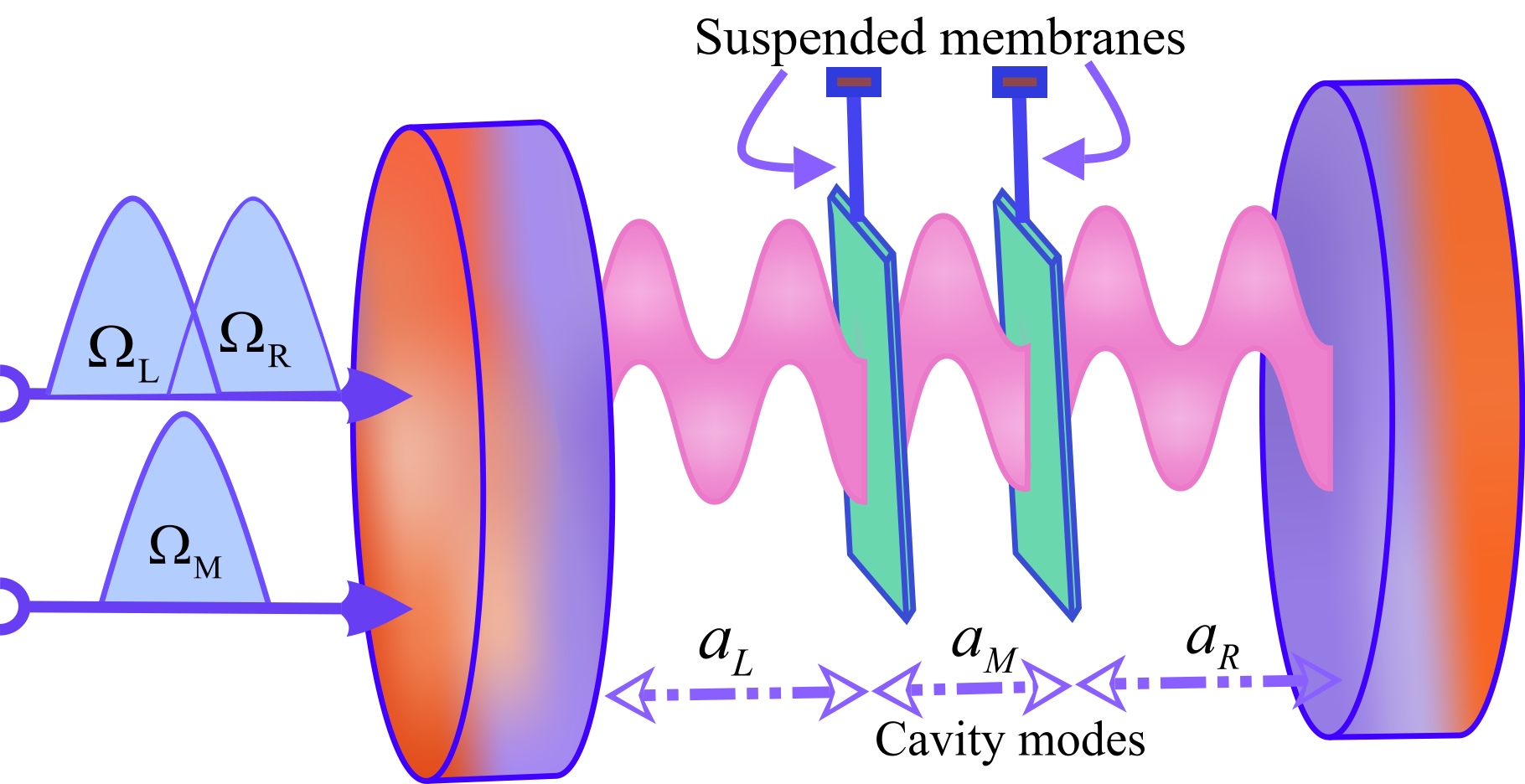}
\caption{Schematic diagram of two membranes inside an optical cavity. }
\label{cavity}
\end{figure}

We consider an optomechanical cavity setup (see Fig. \ref{cavity}) in which two membranes are suspended inside a cavity, dividing it into three subcavities - left (L), middle (M), and right (R). If the membranes are fully reflecting at both of their surfaces, the subcavities can be considered as independent cavities, with a corresponding resonant frequency of $\omega_{n,j}=\frac{n\pi c}{L_j}$ where $n$ is an arbitrary integer, $L_j$ is the length of the $j$th sub-cavity ($j\in L, M, R$), and $c$ is the velocity of light in the vacuum. In case, the $k$th ($k\in 1,2$) membrane is partially transmitting, one can have a tunnelling between the two subcavities on either side of the membrane with a rate $J_k$.  Therefore, the Hamiltonian of the system can be written as follows:
\begin{eqnarray}
\label{hamil0}H_{\rm ac} & = & H_0+H_I+H_{J},\\
H_{0} &=& \sum_{j\in {\rm L,M,R}} \left(\omega_{cj} {a_{j}^\dag} a_{j}\right)+ \sum_{k\in {\rm 1,2}} \left(\omega_{mk} {b_{k}^\dag} b_{k}\right);\nonumber\\
H_I&=&-g_{1} \left( a_{L}^\dag a_{L}-a_{M}^\dag a_{M}\right)\left( b_{1}+b_{1}^\dag\right)\nonumber \\
 & & -g_{2} \left( a_{M}^\dag a_{M}-a_{R}^\dag a_{R}\right)\left( b_{2}+b_{2}^\dag\right),\nonumber\\
H_J & = &-J_{1}\left( a_{L}^\dag a_{M}+ {\rm h.c.}\right)-J_{2}\left( a_{R}^\dag a_{M}+ {\rm h.c.}\right),\nonumber
 \end{eqnarray}
where $H_{0}$ represents the unperturbed Hamiltonian, $a_L$, $a_M$, and $a_R$ are the annihilation operators for the cavity modes of the left, middle, and the right part with corresponding frequencies $\omega_{cj}$ ($j\in L, M, R$) respectively, and $b_{1}$ and $b_{2}$ are the annihilation operators of the mechanical modes with respective frequency $\omega_{m1}$ and $\omega_{m2}$. Further, $g_1$ ($g_2$) determines the optomechanical coupling for first (second) membrane with left (right) and middle mode,
while, $J_1$ ($J_2$) represents the transmission coefficient between the left (right) and middle cavity mode through first (second) membrane.  Note that to obtain a coupling, that is linear in the displacement quadrature $X\equiv (b+b^\dag)$ in a membrane-in-the-middle set up as considered here in $H_I$, one needs to choose $J_{k}=0.5\omega_{mk}$ \cite{law,Ludwig2012,Komar}.

We further drive the $j$th ($j\in {\rm L, M, R}$) cavity mode with the laser field of Rabi frequency $\Omega_j$. It must be borne in mind that these cavity modes are orthogonal to each other, and therefore the laser fields that drive these modes must have orthogonal polarizations to avoid any cross talk. One possible way of doing it may be to send the pulses $\Omega_L$ and $\Omega_R$ (with orthogonal polarizations, and with the same angular frequency $\omega_l$) along the cavity axis to drive the two modes $a_L$ and $a_R$ (see, e.g., \cite{Dong(2012)}), while a linearly polarized pulse $\Omega_M$ may be applied with a different frequency $\omega'_l$ \cite{woolley2014}. This leads to the following Hamiltonian of the system in addition to (\ref{hamil0}):
\begin{equation}
H_{\rm p} = \sum_{j\in {\rm L,R}}\Omega_j \left( a_{j}^\dag e^{-i\omega_{l}t}+{\rm h.c.}\right) + \Omega_M\left( a_{M}^\dag e^{-i\omega'_{l}t}+{\rm h.c.}\right)\;.
\end{equation}
%

In the rotating frame of laser frequencies, the total Hamiltonian $H_{\rm ac}+H_{\rm p}$ takes the following form:
\begin{eqnarray}\label{eq:H}
 H & =&\sum_{j\in {\rm L,M,R}} \left(\Delta_{j} {a_{j}^\dag} a_{j}\right)+ \sum_{k\in {\rm 1,2}} \left(\omega_{mk} {b_{k}^\dag} b_{k}\right)\nonumber \\
 & &-J_{1}\left( a_{L}^\dag a_{M}+ {\rm h.c.}\right)-J_{2}\left( a_{R}^\dag a_{M}+ {\rm h.c.}\right)\nonumber \\
 & &-g_{1} \left( a_{L}^\dag a_{L}-a_{M}^\dag a_{M}\right)\left( b_{1}+b_{1}^\dag\right)\nonumber \\
 & & -g_{2} \left( a_{M}^\dag a_{M}-a_{R}^\dag a_{R}\right)\left( b_{2}+b_{2}^\dag\right)\nonumber \\
 & & +\sum_{j\in {\rm L,M,R}} \Omega_j\left( a_{j}^\dag+{\rm h.c.}\right)\;,
\end{eqnarray}
where $\Delta_{j} =\omega_{cj}- \omega_{l}$ $(j\in L, R)$ are the detunings of the cavity modes $a_L$ and $a_R$ with the driving field, while $\Delta_M=\omega_{cM}-\omega'_l$ is that of the mode $a_M$.
Using the above Hamiltonian and the input-output formalism \cite{Wall(2008)}, we next obtain the following set of Langevin's equations for the relevant operators:
\begin{eqnarray}\label{eq:OLE1}
\dot{a_{L}} &= & -\left(\frac{\gamma_{L}}{2}+i\Delta_{L}\right) a_{L} + i J_{1} a_{M}\nonumber\\
& & +ig_1 a_{L}\left(b_{1}+ b_{1}^{\dag}\right)-i\Omega_{L}-\sqrt{\gamma_{L}} a_{L}^{in},\\
\dot{a_{R}} &= & -\left(\frac{\gamma_{R}}{2}+i\Delta_{R}\right)a_{R} + i J_{2}  a_{M}\nonumber\\
& & -ig_2a_{R}\left(b_{2}+ b_{2}^{\dag}\right)-i\Omega_{R}- \sqrt{\gamma_{R}} a_{R}^{in},\\
\dot{ a_{M}} &= & -\left(\frac{\gamma_{M}}{2}+i\Delta_{M}\right) a_{M} + i J_{2}  a_{R}+ i J_{1}  a_{L}\nonumber\\
& & -ig_1 a_{M}\left(b_{1}+ b_{1}^{\dag}\right)+ig_2 a_{M}\left(b_{2}+ b_{2}^{\dag}\right)-i\Omega_{M}\nonumber\\
& & +\sqrt{\gamma_{M}} a_{M}^{in},\\
\dot{ b_{1}}& = & -\left(\frac{\gamma_{m1}}{2}+i\omega_{m1}\right) b_{1}+ ig_{1}\left(a_{L}^{\dag} a_{L}-a_{M}^{\dag}a_{M}\right)\nonumber\\
& &+\sqrt{\gamma_{m}} b_{1}^{in},\\\label{eq:OLE2}
\dot{ b_{2}}& = & -\left(\frac{\gamma_{m2}}{2}+i\omega_{m2}\right) b_{2} -ig_{2}\left(a_{R}^{\dag}a_{R}-a_{M}^{\dag}a_{M}\right)\nonumber\\
& &+\sqrt{\gamma_{m}} b_{2}^{in},
\end{eqnarray}
where $\gamma_j$ is the decay rate of the $j$th mode of the cavity and $\gamma_{mk}$ is the dissipation rate of the $k$th membrane. The corresponding noise operators $a_{j}^{in}$ ($j\in L, M, R$) and  $b_{k}^{in}$ ($k\in 1,2$) satisfy the following correlations \cite{Wall(2008)}:
\begin{eqnarray}
 \left\langle a_{j}^{in}(t) a_{j}^{\dag in}(t')\right\rangle &= &  \delta(t-t'),\nonumber\\
 \left\langle\ a_{j}^{\dag in}(t) a_{j}^{in}(t')\right\rangle &= &   0 ,\nonumber\\
 \left\langle b_{k}^{in}(t) b_{k}^{in \dag}(t')\right\rangle & = &(\bar{n}_{th}+1)\delta(t-t'),\nonumber\\
  \left\langle b_{k}^{in\dag}(t) b_{k}^{in }(t')\right\rangle & = &(\bar{n}_{th})\delta(t-t'),
\end{eqnarray}
where $\bar{n}_{th}=\left\{ \exp\left[\hbar\omega_{mk}/(k_{B}T)\right]-1\right\}^{-1}$ is the mean thermal excitation number in the bath, interacting with the mechanical oscillator with frequency $\omega_{mk}$ at an equilibrium temperature $T$ and $k_{B}$ is the Boltzmann constant.

\subsection{Derivation of effective Hamiltonian}\label{SAE}
In order to study the dynamics of the cavity mode and the membranes, we use the standard linearization procedure \cite{Aspelmeyer}, in which one expands all the bosonic operators as a sum of the average values and the zero-mean fluctuation, as follows: $a_j\rightarrow \alpha_j + \delta a_j$, $b_k\rightarrow \beta_k + \delta b_k$. Here $\alpha_j$ and $\beta_k$ are in general complex and denote the steady state values of the respective annihilation operators of the cavity and  membrane modes. Applying this transformation to Eqs. (\ref{eq:OLE1}-\ref{eq:OLE2}), we obtain the following equations for the average of the operators:
\begin{eqnarray}
\dot{\alpha_{L}} &= & -\left(\frac{\gamma_{L}}{2}+i\Delta_{L}'\right)\alpha_{L} + i J_{1} \alpha_{M}-i\Omega_{L}\\
\dot{\alpha_{R}} &= & -\left(\frac{\gamma_{R}}{2}+i\Delta_{R}'\right)\alpha_{R} + i J_{2} \alpha_{M} -i\Omega_{R}\\
\dot{ \alpha_{M}} &= & -\left(\frac{\gamma_{M}}{2}+i\Delta_{M}'\right) \alpha_{M} + i J_{1}  \alpha_{L}+ i J_{2}  a_{R}\nonumber \\
 & &-i\Omega_{M}\\
\dot{ \beta_{1}}& = & -\left(\frac{\gamma_{m1}}{2}+i\omega_{m1}\right) \beta_{1}+ ig_{1}\left(|\alpha_{L}|^2-|\alpha_{M}|^2\right)\\
\dot{\beta_{2}}& = & -\left(\frac{\gamma_{m2}}{2}+i\omega_{m2}\right)\beta_{2} -ig_{2}\left(|\alpha_{R}|^2-|\alpha_{M}|^2\right)\;,
\end{eqnarray}
where $\Delta'_L=\Delta_L-g_{1}\left( \beta_{1}+\beta_{1}^*\right)$ , $\Delta'_R=\Delta_R+g_{2} \left( \beta_{2}+\beta_{2}^*\right)$ and $\Delta'_M=\Delta_M+g_{1}\left( \beta_{1}+\beta_{1}^*\right)-g_{2}\left( \beta_{2}+\beta_{2}^*\right)$  represent the modified detunings of the respective cavity mode.
The steady state solutions for the above equations can be found by taking the time-derivatives equal to zero, as
\begin{eqnarray}\label{alpha1}
\alpha_{L} &=& \frac{\Omega_{L}-J_1\alpha_{M}}{-\Delta_{L}'+i\frac{\gamma_{L}}{2}},\\
\label{alpha2}\alpha_{R} &=& \frac{\Omega_{R}-J_2\alpha_{M}}{-\Delta_{R}'+i\frac{\gamma_{R}}{2}},\\
\label{alpha3}\alpha_{M} &=& \frac{\Omega_{M}-J_1\alpha_{L}-J_2\alpha_{R}}{-\Delta_{M}'+i\frac{\gamma_{M}}{2}}
\end{eqnarray}
Similarly, the Langevin equations for the fluctuations can be obtained  using Eqs. (\ref{eq:OLE1}-\ref{eq:OLE2}), as follows:
\begin{eqnarray}\label{eq:FE1}
\dot{\delta a_{L}} &= & -\left(\frac{\gamma_{L}}{2}+i\Delta'_{L}\right)\delta a_{L} + i J_{1} \delta a_{M}+ig_1\alpha_{L}\left(\delta b_{1}+ \delta b_{1}^{\dag}\right)\nonumber\\
& & + \sqrt{\gamma_{L}} \delta a_{L}^{in},\\
\dot{\delta a_{R}} &= & -\left(\frac{\gamma_{R}}{2}+i\Delta'_{R}\right)\delta a_{R} + i J_{2} \delta a_{M}+ig_2\alpha_{L}\left(\delta b_{2}+ \delta b_{2}^{\dag}\right)\nonumber\\
& & + \sqrt{\gamma_{R}} \delta a_{R}^{in},\\
\dot{\delta a_{M}} &= & -\left(\frac{\gamma_{M}}{2}+i\Delta'_{M}\right)\delta a_{M} + i J_{1} \delta a_{L}+ i J_{2} \delta a_{R}+ \sqrt{\gamma_{M}} \delta a_{M}^{in}\nonumber\\
& & -ig_1\alpha_{M}\left(\delta b_{1}+ \delta b_{1}^\dag\right)+ig_2\alpha_{M}\left(\delta b_{2}+ \delta b_{2}^{\dag}\right),\\
\dot{\delta b_{1}}& = & -\left(\frac{\gamma_{m1}}{2}+i\omega_{m1}\right)\delta b_{1} + ig_{1}\left[\alpha_{L}\left(\delta a_{L}+ \delta a_{L}^{\dag}\right)\right.\nonumber\\
&&\left. -\alpha_{M}\left(\delta a_{M}+\delta{a_{M}^{\dag}}\right)\right]+\sqrt{\gamma_{m1}} \delta b_{1}^{in},\\\label{eq:FE2}
\dot{\delta b_{2}}& = & -\left(\frac{\gamma_{m2}}{2}+i\omega_{m2}\right)\delta b_{2} -ig_{2}\left[\alpha_{R}\left(\delta a_{R}+ \delta a_{R}^{\dag}\right)\right.\nonumber\\
& &\left.-\alpha_{M}\left(\delta a_{M}+\delta{a_{M}^{\dag}}\right)\right] +\sqrt{\gamma_{m2}} \delta b_{2}^{in},
\end{eqnarray}
 The above equations can be derived from the following linearized Hamiltonian:
  \begin{eqnarray}\label{eq:FH}
 H & =&\sum_{j \in {\rm L,M,R}} \left(\Delta'_{j} {\delta a_{j}^\dag} \delta a_{j}\right)+ \sum_{k\in {\rm 1,2}} \left(\omega_{mk} {\delta b_{k}^\dag} \delta b_{k}\right)\nonumber \\
 & &-J_{1}\left( \delta a_{L}^\dag \delta a_{M}+ {\rm h.c.}\right)-J_{2}\left( \delta a_{R}^\dag \delta a_{M}+ {\rm h.c.}\right)\nonumber \\
 & &-g_{1} \left( \alpha^*_{L} \delta a_{L}+\alpha_{L} \delta a_{L}^\dag\right)\left( \delta b_{1}+\delta b_{1}^\dag\right)\nonumber \\
 & &+g_{2} \left( \alpha^*_{R} \delta a_{R}+\alpha_{R} \delta a_{R}^\dag\right)\left( \delta b_{2}+\delta b_{2}^\dag\right)\nonumber \\
 & &+g_{1} \left( \alpha^*_{M} \delta a_{M}+\alpha_{M} \delta a_{M}^\dag\right)\left( \delta b_{1}+\delta b_{1}^\dag\right)\nonumber \\
 & &-g_{2} \left( \alpha^*_{M} \delta a_{M}+\alpha_{M} \delta a_{M}^\dag\right)\left( \delta b_{2}+\delta b_{2}^\dag\right)\;.
 \end{eqnarray}

We choose the laser frequencies in the red sideband region so that
\begin{eqnarray}\label{rwacond}
\Delta'_j=\omega_{mk}\;;\;\;j\in L,M,R\;\;;\;\; k\in 1,2\;.
\end{eqnarray}
This corresponds to the following relation between the cavity mode frequencies:
\begin{equation}
\omega_{cL}-\omega_{cM}=(\omega_l-\omega'_l)-3g_2(\beta_2+\beta_2^*)\;,
\end{equation}
while $g_1(\beta_1+\beta_1^*)=-g_2(\beta_2+\beta_2^*)$. This further requires $g_1$ and $g_2$ to be out-of-phase, for positive real values of $\beta_{1,2}$, and  this is achievable in the present configuration [see Eq. (56), \cite{law}]. The above choice of sideband clearly allows us to drive three modes of the cavity with three different pulses of suitable polarization, frequencies, and time-dependences, without any possibility of the cross-talk. The separation between the frequencies of the two sub-cavity modes are chosen to be much larger than their respective line-widths $\gamma_{mk}$. 

In the interaction picture with respect to the unperturbed part of the above Hamiltonian
\begin{eqnarray}
 H_{0} & =&\sum_{j\in {\rm L,M,R}} \left(\Delta'_{j} {\delta a_{j}^\dag}\delta a_{j}\right)+ \sum_{k\in {\rm 1,2}} \left(\omega_{mk} {\delta b_{k}^\dag} \delta b_{k}\right)\;,
\end{eqnarray}
the condition (\ref{rwacond}) and the weak-coupling condition \cite{Tian2010} 
$\omega_{mk} \gg |g_{1}\alpha_L|, |g_{2}\alpha_R|, |g_{1}\alpha_M|, |g_{2}\alpha_M|  $
allow us to take the rotating wave approximation and to obtain the following final form of the effective Hamiltonian:
\begin{eqnarray}\label{eq:IH}
 H & =&-J_{1}\left( \delta a_{L}^\dag \delta a_{M}+ {\rm h.c.}\right)-J_{2}\left( \delta a_{R}^\dag \delta a_{M}+ {\rm h.c.}\right)\nonumber \\
 & &-g_{1} \left( \alpha^*_{L} \delta b_{1}^{\dag} \delta a_{L}-\alpha^*_{M}\delta b_{1}^\dag \delta a_{M}+{\rm h.c.}\right)\nonumber \\
 & &+g_{2} \left( \alpha^*_{R}\delta b_{2}^\dag \delta a_{R}-\alpha^*_{M}\delta b_{2}^\dag \delta a_{M}+{\rm h.c.}\right)\;.
 \end{eqnarray}

The Langevin equations for the operator fluctuations can then be obtained, using the above Hamiltonian, as follows:
\begin{eqnarray}\label{IFE1}
\dot{\delta a_{L}} &= &   i J_{1} \delta a_{M}+ig_1\alpha_{L}\delta b_{1}-\frac{\gamma_{L}}{2}\delta a_{L}+ \sqrt{\gamma_{L}} \delta a_{L}^{in}\\
\dot{\delta a_{R}} &= & i J_{2} \delta a_{M}-ig_2\alpha_{R}\delta b_{2}-\frac{\gamma_{R}}{2}\delta a_{R} + \sqrt{\gamma_{R}} \delta a_{R}^{in}\\
\dot{\delta a_{M}} &= &  i J_{1} \delta a_{L}+ i J_{2} \delta a_{R}-ig_1\alpha_{M}\delta b_{1}+ig_2\alpha_{M}\delta b_{2}\nonumber\\
& & -\frac{\gamma_{M}}{2}\delta a_{M}+ \sqrt{\gamma_{M}}\delta  a_{M}^{in}\\
\dot{\delta b_{1}}& = & ig_{1}\left(\alpha_{L}\delta a_{L}-\alpha_{M}\delta a_{M}\right)-\frac{\gamma_{m1}}{2}\delta b_{1}\nonumber\\
& &+\sqrt{\gamma_{m1}} \delta b_{1}^{in}\\\label{IFE2}
\dot{\delta b_{2}}& = & -ig_{2}\left(\alpha_{R}\delta a_{R}-\alpha_{M}\delta a_{M}\right)-\frac{\gamma_{m2}}{2}\delta b_{2}\nonumber\\
& & +\sqrt{\gamma_{m2}} \delta b_{2}^{in}.
\end{eqnarray}

\section{Adiabatic transfer} \label{EAEV}
The above set of equations (\ref{IFE1}-\ref{IFE2}) can be written in a matrix form as
        $ i\dot{F}=MF$, where
 \begin{eqnarray}
       F= \begin{bmatrix}
        \delta a_{L}&\delta a_{M}&\delta a_{R}&\delta b_{1}&\delta b_{2}
        \end{bmatrix}^T,
\end{eqnarray}
with $T$ representing the transpose of the matrix and
\begin{eqnarray}
M=
 \begin{bmatrix}
   -i\frac{\gamma_{L}}{2} & -J_{1} & 0 & -g_1\alpha_{L}\ & 0 \\
   -J_1   & -i\frac{\gamma_{M}}{2} & -J_2 & g_{1}\alpha_{M} & -g_{2}\alpha_{M} \\
    0   & -J_2 & -i\frac{\gamma_{R}}{2} & 0 & g_2\alpha_{R} \\
   -g_1\alpha_{L} & g_1\alpha_{M} & 0 & -i\frac{\gamma_{m1}}{2} & 0 \\
   0&-g_2\alpha_{M} & g_2\alpha_{R} & 0 & -i\frac{\gamma_{m2}}{2}
\end{bmatrix}.
\end{eqnarray}
We find that the above matrix $M$ exhibits a zero eigenvalue, in absence of the decay terms, with the corresponding eigenmode:
\begin{eqnarray}\label{dark}
\psi_{D}=\left(2g_1g_2\alpha_{L}\alpha_{R}\right)\hat{\delta a_M}-\left(g_2\alpha_{R}\right)\hat{\delta b_1}+\left(g_1\alpha_{L}\right)\hat{\delta b_2},\nonumber\\
 \end{eqnarray}
where we have considered $\alpha_L$ and $\alpha_R$ to be real [in absence of decay terms, see Eqs. (\ref{alpha1}) and (\ref{alpha2})] and $\alpha_M=0$.
The mode (\ref{dark}) represents a zero-eigenvalue adiabatic mode. Comparing this eigenmode with the state (\ref{adiab}) for STIRAP, we find that it is possible to transfer the excitation of the mode $b_1$ to the mode $b_2$, by suitably choosing the time-dependence of $\alpha_{L}$ and $\alpha_{R}$.  Interestingly, the above mode is spanned over the mode $a_M$ also. This suggests that the time-evolution should be fast enough to avoid the decay of the cavity mode $a_M$.  We emphasize that, to have $\alpha_M=0$, the middle mode also needs to be driven by another field $\Omega_M$, given by
\begin{equation}\label{omegam_pulse}
\Omega_{M}= J_{1}\alpha_{L} +J_{2} \alpha_{R}\;,
\end{equation}
as clear from the Eq. (\ref{alpha3}).

From the steady state expressions (\ref{alpha1}) and (\ref{alpha2}), it is obvious that the time-dependence of $\alpha_L$ and $\alpha_R$ are effectively governed by the driving fields $\Omega_L$ and $\Omega_R$.
We choose these fields with the following Gaussian envelope in time-domain:
 \begin{eqnarray}\label{G1}
 \Omega_L(t) &=& A\exp\left[-(t-\tau)^2)/T^2\right],\\
 \label{G2}\Omega_R(t) &=& A\exp\left[-(t+\tau)^2/T^2\right],\\
 \label{G3}\Omega_M(t) &=& J_1\frac{\Omega_L}{-\Delta_{L}'}+J_2\frac{\Omega_R}{-\Delta_{R}'}\;.
\end{eqnarray}
Here $A$ represents the amplitude of the Gaussian pulses, $T$ is the width of the pulses, and $2\tau$ represents the pulse delay between the pulses. Note that such a time-delay between the pulses can be obtained by introducing path-difference, as routinely done in standard optical experiments. The time-dependent optomechanical coupling can then be written, using Eqs. (\ref{alpha1}-\ref{alpha2}) and for negligible decay rates, as
\begin{eqnarray}\label{g1aL}
g_{1}\alpha_{L}& = &  \frac{g_{1}A}{-\Delta^{'}_{L}}\exp{\left[-(t-\tau)^2/T^2\right]}\\
\label{g2aR}g_{2}\alpha_{R} & = &  \frac{g_{2}A}{-\Delta^{'}_{R}}\exp{\left[-(t+\tau)^2/T^2\right]}\;.
\end{eqnarray}
This represents a delayed and counter-intuitive pulse sequence to transfer the fluctuation excitation from the $b_1$ mode to the $b_2$ mode, through the evolution of the eigenmode $\psi_D$, akin to STIRAP.

\begin{figure}
\centering
\includegraphics[width=0.4\textwidth]{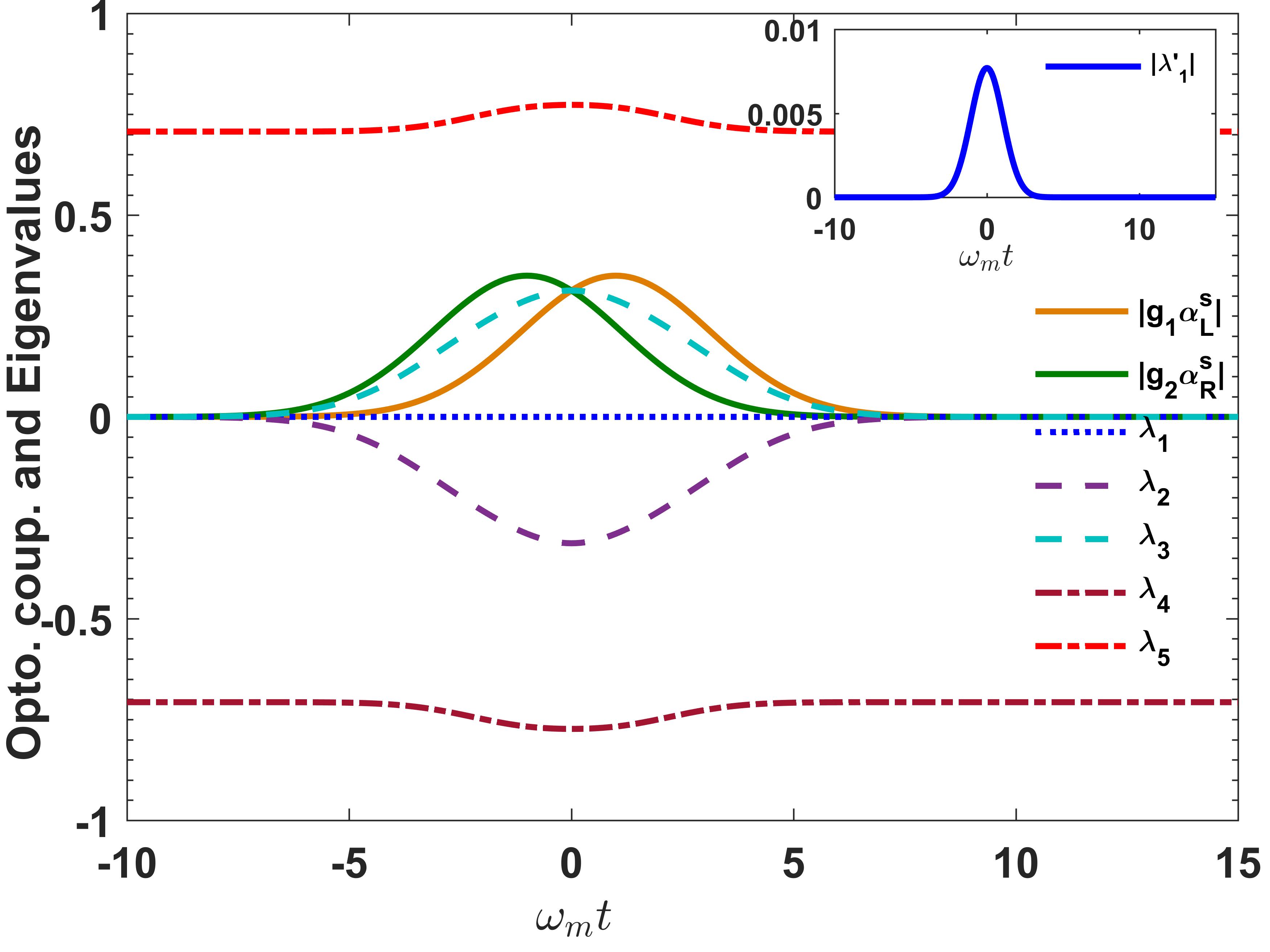}
\caption{Time variation of the eigenvalues (\ref{EV}) of the matrix $M$ and the time-dependent optomechanical couplings (\ref{g1aL}) and (\ref{g2aR}). The inset displays the time-variation of the first-order shift of the eigenvalue $\lambda_1=0$, in presence of decay. All the parameters are normalized with respect to 1 MHz. Parameters are $A=350$, $\Delta'_j=\omega_{mk}=1$, $g_k=0.001$, $J_k=1/2$, $T=3$, $\tau=1$, $\gamma_{j}=0$, $\gamma_{mk}=0$, for all $j\in L,M,R$ and $k\in 1,2$.}\label{eigenvalues}
\end{figure}

Note that the other eigenvalues of $M$ are given by
\begin{eqnarray}\label{EV}
\lambda_{2}&=&-\lambda_{3} = -\frac{1}{2} \sqrt{ \left(\alpha_{0} - \sqrt{\beta_{0}}\right)},\nonumber\\
\lambda_{4}&=&-\lambda_{5} = -\frac{1}{2} \sqrt{\left(\alpha_{0} + \sqrt{\beta_{0}}\right)}
\end{eqnarray}
where\begin{eqnarray}
\alpha_{0}& = & 1 + 2 \alpha_{L}^{2} g_{1}^2 + 2  \alpha_{R}^{2}g_{2}^2\\
\beta_{0} & = & 1 + 4  \alpha_{L}^{4}g_{1}^4 - 8
 \alpha_{L}^{2} \alpha_{R}^{2} g_{1}^2 g_{2}^2 + 4\alpha_{R}^{4}  g_{2}^4\;.
\end{eqnarray}
 We show in Fig. \ref{eigenvalues}, how these eigenvalues $\lambda_i$ ($i\in 1,\cdots,5$) vary with time under the action of these pulses. We find that the gap between the two eigenvalues become larger during the maximum overlap of the two pulses. This gap reflects that the system would remain confined in the zero-eigenvalue eigenstate, as there is no level-crossing during evolution, ensuring the adiabaticity of the process.
\begin{figure}
\centering
\includegraphics[width=0.43\textwidth]{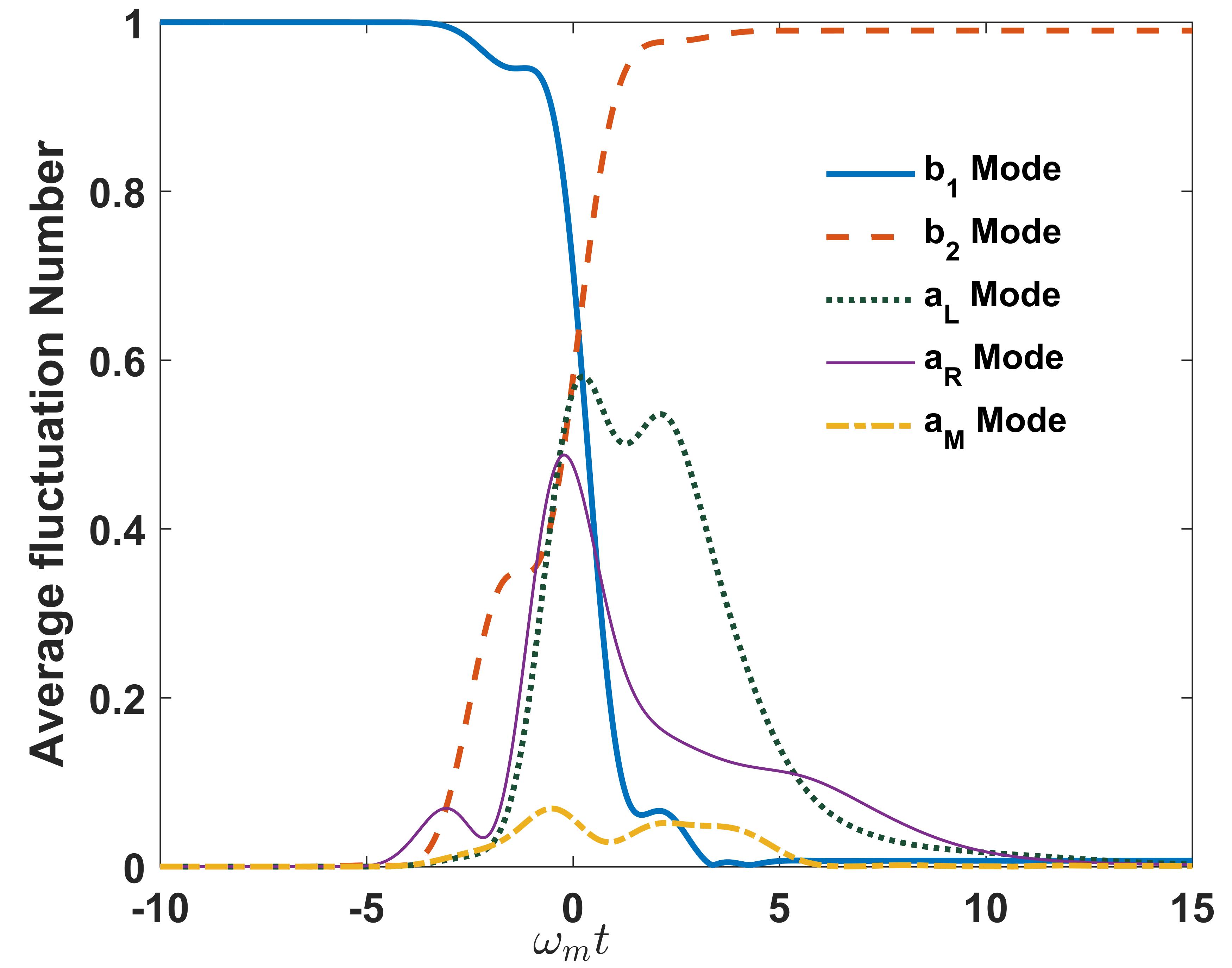}
\caption{Time-evolution of the average excitation fluctuations of the two membranes and the three sub-cavity modes. We have chosen initial average excitation fluctuations of the first membrane as $\langle\delta b_{1}^\dag \delta b_{1}\rangle =1$. We have chosen $\gamma_{j}=0.4$ and $\gamma_{mk}=0.0001$. All the other parameters are the same as in Fig. \ref{eigenvalues}.}\label{g1}
\end{figure}

We choose the pulse sequence  (\ref{G1})-(\ref{G3})  to drive the cavity modes. The pulse $\Omega_R(t)$ is applied first on the right cavity R, thereby increasing the average photon number in that mode and effectively the radiation pressure on the second membrane. This leads to a driven oscillation of the right membrane, with an enhanced phonon excitation. Next, the left cavity L is driven by a pulse $\Omega_L(t)$, partially overlapping with the $\Omega_R(t)$. This causes the phonon population sweep at resonance, in analogy of STIRAP. In this way, the excitation fluctuation of the left membrane is adiabatically  transferred to the other  mediated by the middle sub-cavity mode $a_M$. 

To verify the adiabatic transfer of the excitation, as discussed above, we next solve numerically the Langevin equations (\ref{IFE1}-\ref{IFE2}) and investigate the time-evolution of the average phonon number fluctuation $\langle\delta b_1^\dag\delta b_1\rangle$ and $\langle\delta b_2^\dag \delta b_2\rangle$. For an initial average phonon number fluctuation $\langle \delta b_{1}^\dag \delta b_{1}\rangle =1 $ in mechanical mode $b_{1}$,  we display in Fig. \ref{g1} the transfer of excitation to the $b_2$ mode. It is observed that for a weak optomechanical coupling ($g_{1}=g_{2}=0.001$), complete population transfer would require large amplitude pulses. Larger coupling strength (e.g., $g_{1}=g_{2}=0.01$) would relax the requirement of such large amplitudes for complete excitation transfer.  It also populates the middle mode $a_M$  through tunnelling via the two mirrors, the excitation $\langle\delta a_M^\dag\delta a_M\rangle$ of which however decays to zero at the steady state. 

We emphasize here that to maintain the adiabaticity in the process, the pulse area should be large so that  the system remains in the zero-eigenvalue adiabatic state for all the times and the transition probability to any other adiabatic state remains negligible. Moreover, the decay rates of the membranes are assumed to be much smaller than their fundamental frequencies such that there is no significant decay of the phonons during the transfer process. If we consider the damping terms of our system as a perturbation, we can obtain the deviation in the eigenvalue $\lambda_1$ from zero, using the first order perturbation theory. The matrix $M$ can be divided into two parts: one $M_{nodec}$ without the decay term and the other $M_{decay}$ that includes elements involving only the decay rates, i.e., $M=M_{nodec}+M_{decay}$, where $M_{decay}$ can be written as
\begin{equation}\label{decayeigen}
M_{decay}=-i\;{\rm diag}\left[\frac{\gamma_L}{2},\frac{\gamma_M}{2},\frac{\gamma_R}{2},\frac{\gamma_{m1}}{2},\frac{\gamma_{m2}}{2}\right]\;.
\end{equation}
So the first-order correction of the eigenvalue can be obtained as
\begin{eqnarray}\label{perturb}
\lambda'_1 &=&\psi_D^TM_{decay}\psi_D\nonumber\\
&=&-2i\gamma_{M}(g_1g_2\alpha_{L}\alpha_{R})^2-i\frac{\gamma_{m1}}{2}({g_2\alpha_{R}})^{2} -i\frac{\gamma_{m2}}{2}(g_{1}\alpha_{L})^{2}\;,\nonumber\\
\end{eqnarray}
where the zero-eigenvalue eigenmode $\psi_D$ is considered in its matrix form. We display the temporal variation of this eigenvalue $|\lambda'_1|$ in the inset of the Fig. \ref{eigenvalues}. Clearly this does not deviate much from the value zero. This suggests that the moderate values of decay rates of the sub-cavity modes and the membranes do not affect much the adiabatic transfer. This is further verified in the Fig. \ref{g1}, which shows that the transfer is nearly complete in presence of the decay of all the modes. In fact, negative imaginary shift of the eigenvalue would otherwise lead to the decay of the eigenmode 
$\psi_D$; however, the magnitude of this decay rate $|\lambda'_1|$ remains negligibly small during the adiabatic transfer.


\section{conclusion}           \label{con}
In summary, we have described a scheme in an optomechanical set up to adiabatically transfer the average fluctuation of excitation from one membrane to the other, both being suspended inside an optical cavity. These membrane, while suitably placed, divide the entire cavity into three sub-cavities - L, M, and R. The corresponding optical modes couple with each other  via tunnelling through the membranes and to the membranes through radiation pressure force. We propose to drive the left and right modes with time-dependent pulses in counter-intuitive sequences such that a zero-eigenvalue adiabatic mode is obtained, analogous to that obtained for STIRAP.  Thereby, the effective couplings between the membranes and the optical modes become time-dependent and facilitate transfer of excitation fluctuation through this adiabatic mode. In addition, a third pulse, as a suitable superposition of these two delayed pulses, is employed to drive the middle cavity mode so as to avoid its excitation. We have analyzed  the adiabatic features of the transfer, through the time-evolution of the relevant eigenvalues. We have further investigated, both analytically and numerically,  the robustness of this evolution against the decays of the membranes and the cavity modes. We emphasize that the exchange of the energy fluctuations between two mechanical systems may pose as a possible quantum communication protocol. The slow decay of the mechanical oscillator may facilitate the information storage as well.   

\end{document}